\documentclass[11pt]{article}
\usepackage{moriond,epsfig}

\def\APP{{\em Astropart. Phys.} }

\def \icrc{[Pierre Auger Collaboration], to appear in {\em Proc. 30th ICRC } (2007).}
\def \icrcold{[Pierre Auger Collaboration], {\em Proc. 29th ICRC} {\bf 7} (2005)}
\begin{document}
\vspace*{4cm}
\title{PIERRE AUGER OBSERVATORY STATUS AND RESULTS}

\author{ V. VAN ELEWYCK \\
for the Pierre Auger Collaboration}

\address{Institut de Physique Nucl\'eaire d'Orsay, Universit\'e de Paris Sud  \& CNRS-IN2P3 \\
15, rue G. Clemenceau, 91406 Orsay Cedex, France\\}

\maketitle\abstracts{The Pierre Auger Observatory, a hybrid detector
for the study of ultra-high energy cosmic rays (UHECRs), is now
approaching completion. After describing Auger present status and
performance, with an emphasis on the advantages provided by the
combination of two different detection techniques, this contribution
presents a brief panorama of the first scientific results achieved
and of their impact on our knowledge of the UHECRs' origin and
composition.}

\section{Introduction}
Despite the important progress achieved in cosmic ray physics during
the last decades, fundamental questions about the nature and origin
of the ultra-high energy cosmic rays (UHECR) are still unanswered.
Contradictory results~\cite{agasa,HiRes} have been reported about
the presence of the expected GZK cutoff in the cosmic ray spectrum
at energies around $5\times 10^{19}$ eV; and the identification of
possible acceleration sites still awaits the observation of an
unambiguous correlation of UHECR with astrophysical objects
(see~\cite{phys} for recent reviews on these
issues). 
The Pierre Auger Observatory~\cite{opa}  is expected to shed some
light on these longstanding questions. 
  Its hybrid design, combining a surface detector (SD) and a fluorescence detector (FD), makes it
sensitive to different - and complementary - observables of the
extensive air showers (EAS) related to the primary UHECR properties.
With more than 75\% of the SD stations deployed and all four
fluorescence telescopes operational at the time of writing, the
Auger Southern Site (located in the province of Mendoza, Argentina)
is now nearing completion and has been accumulating high-quality
data at a regularly increasing pace for the past three years. 
After a brief description of the detector and its current
performance in Sec.~\ref{sec:status}, a review of the significant
physics results already produced by Auger concerning the UHECR
energy spectrum (Sec.~\ref{sec:spectrum}), arrival directions
(Sec.~\ref{sec:ARG}) and composition (Sec.~\ref{sec:compo}) will be
presented.

\section{Status and description of the observatory and its dataset}
\label{sec:status}
 The SD is a triangular array of 1600 water tanks
distant 1.5 km from each other, which sample the shower content at
ground. The Cherenkov light emitted by the particles entering the
tank is detected by three photomultipliers and the corresponding
signals are digitized at 40 MHz by Flash Analog-Digital Converters
(FADC). Two local
triggers are used: a simple ``Threshold'' (Th) one, 
 and a ``Time-over-Threshold'' (ToT)
one which requires a lower but more extended signal (at least 12
FADC bins) and is more sensitive to the electromagnetic (EM)
component of the shower. A global trigger (T3) then asks for a
relatively compact configuration of local triggers compatible in
time with the arrival of a shower front. Finally, offline criteria
are applied to reject accidental triggers (``physics trigger'', T4)
and to ensure the reconstructibility of the events (``quality
trigger'',T5) \cite{triggers}. The SD is constantly active and
provides the bulk of data required for high-statistics analysis. Its
detection efficiency is 100 \% above $10^{18.5}$ eV at zenith angles
below $60^\circ$. The angular
accuracy on the arrival direction is determined 
on the basis of an empirical model for the time measurement
uncertainties~\cite{carla}; it depends on the number of hit stations
but is always better than $2^\circ$ for events at $\theta \leq
60^\circ$ .


The SD array is overlooked by four FD sites that measure the
ultraviolet light produced when charged particles in the air shower
excite nitrogen molecules in the atmosphere. Each site features six
Schmidt telescopes that cover a field of view of about $30^\circ
\times 30^\circ$ each. The signal is collected on a 440 pixels
camera and digitalized at a 100 MHz sampling rate. The fluorescence
light emitted by the shower is roughly proportional to the energy
dissipated in the atmosphere. The fluorescence telescopes can be
used only during dark, moonless nights, which reduce their duty
cycle to about $14\%$. The timing and position of the triggering
pixels allow to reconstruct the shower-detector plane with an
accuracy of about $0.3^\circ$, but the uncertainty on the
orientation of the shower axis within that plane is much larger.

Most events seen by the FD also trigger at least one SD station, and
the additional timing information allows to significantly improve
the accuracy both on the reconstructed arrival direction ($\sim
0.5^\circ$) and the position of the core ($\sim 50 \mathrm{m}$).
These {\it hybrid} events amount to about 10\% of the total data
sample; they allow to fully exploit the detector capabilities and
therefore have an important impact on many analysis performed on
Auger data. High-quality events, which independently trigger the FD
and the SD and can be successfully reconstructed by both detectors,
are tagged as {\it golden hybrids}. These events allow the
simultaneous measurement and cross-calibration of different shower
parameters related to the energy and nature of the UHECRs. Such a
strategy allows to extract physical information about their spectrum
and composition while minimizing the dependance in model
assumptions, as will be illustrated in the next sections.




\section{The spectrum of UHECR}
\label{sec:spectrum}

\subsection{General strategy}
\label{subsec:strategy}
 The key ingredients for the determination of the UHECR spectrum are the
accurate determination of the primary energy, which is best achieved
with the fluorescence technique, and a large and easily calculable
exposure, which is provided by the SD. 

 The signals in the triggered
stations are used to reconstruct both the shower core position and
its lateral profile at ground. The parameter S(1000), {\it i.e.} the
signal that would be produced in a tank located at 1000~m from the
shower core, is measured with an accuracy better than 12\% and can
be used as an estimator of the size (and thus energy) of the
shower~\cite{ave}. For a given energy, its value depends on the
zenith angle of the shower as a consequence of geometrical
effects and of the attenuation of the shower in the atmosphere. 
The ``constant intensity cut" method \cite{sommers} is used to
extract the shape of this attenuation curve, $CIC(\theta)$, 
from the data assuming an isotropic flux of
UHECRs. 
The S(1000) is then converted into a reference value taken at the
mean of the zenith angle interval,
$S_{38^\circ} \equiv S(1000)/CIC(\theta)$.

The relation of $S_{38^\circ}$ (or S(1000)) to the primary energy
however significantly depends on the assumptions on the primary
composition and on hadronic models which drive the development of
the shower. This drawback is circumvented by using the golden hybrid
events to calibrate $S_{38^\circ}$ on the energy obtained with the
FD. The information on the shower longitudinal profile provided by
the FD indeed allows an independant, nearly-calorimetric measurement
of the energy of the shower, $E_{FD}$. Dependance in composition and
hadronic models only affects the determination of the {\it
invisible} component,
{\it i.e.} muons and neutrinos, which 
contributes only 4\% of the total uncertainty in the FD energy. More
significant sources of systematics are the uncertainty on the
fluorescence yield and its dependance in the atmospherical
conditions, the absolute calibration of the FD and the energy
reconstruction method. Current estimations \cite{dawson} of the
overall systematics in $E_{FD}$ give about 22\%, while the
statistical uncertainty in the derived energy is smaller than 10\%.

\begin{figure}[t]
\begin{minipage}[t]{0.48\textwidth}
\mbox{}\\
\hfill
\centerline{\includegraphics[width=\textwidth,height=6cm]{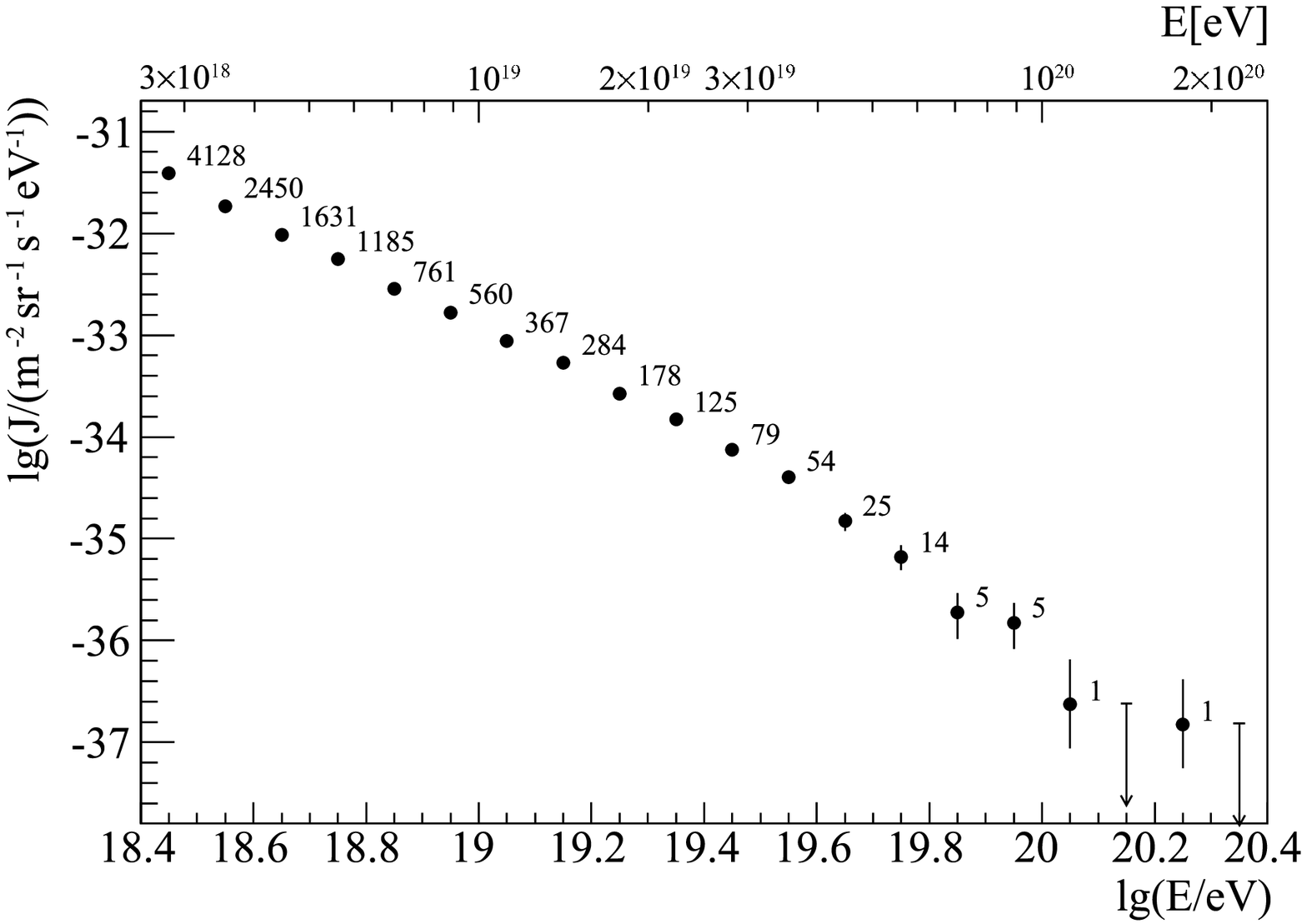}}
\hfill \vspace*{-0.5cm} \caption[spectrum]{\raggedright
\footnotesize {Auger spectrum J in function of the energy (with the
number of events per bin), obtained from the SD dataset as described
in Sec.\ref{subsec:strategy}. Vertical error bars are statistical
only. Stat. and syst. errors in the energy scale are $\approx 6\%$
and $\approx 22\%$ (from~\cite{roth}).}} \label{fig:spectrumSD}
\end{minipage}
\hfill
\begin{minipage}[t]{0.48\textwidth}
\mbox{}\\
\centerline{\includegraphics[width=\textwidth,height=6cm]{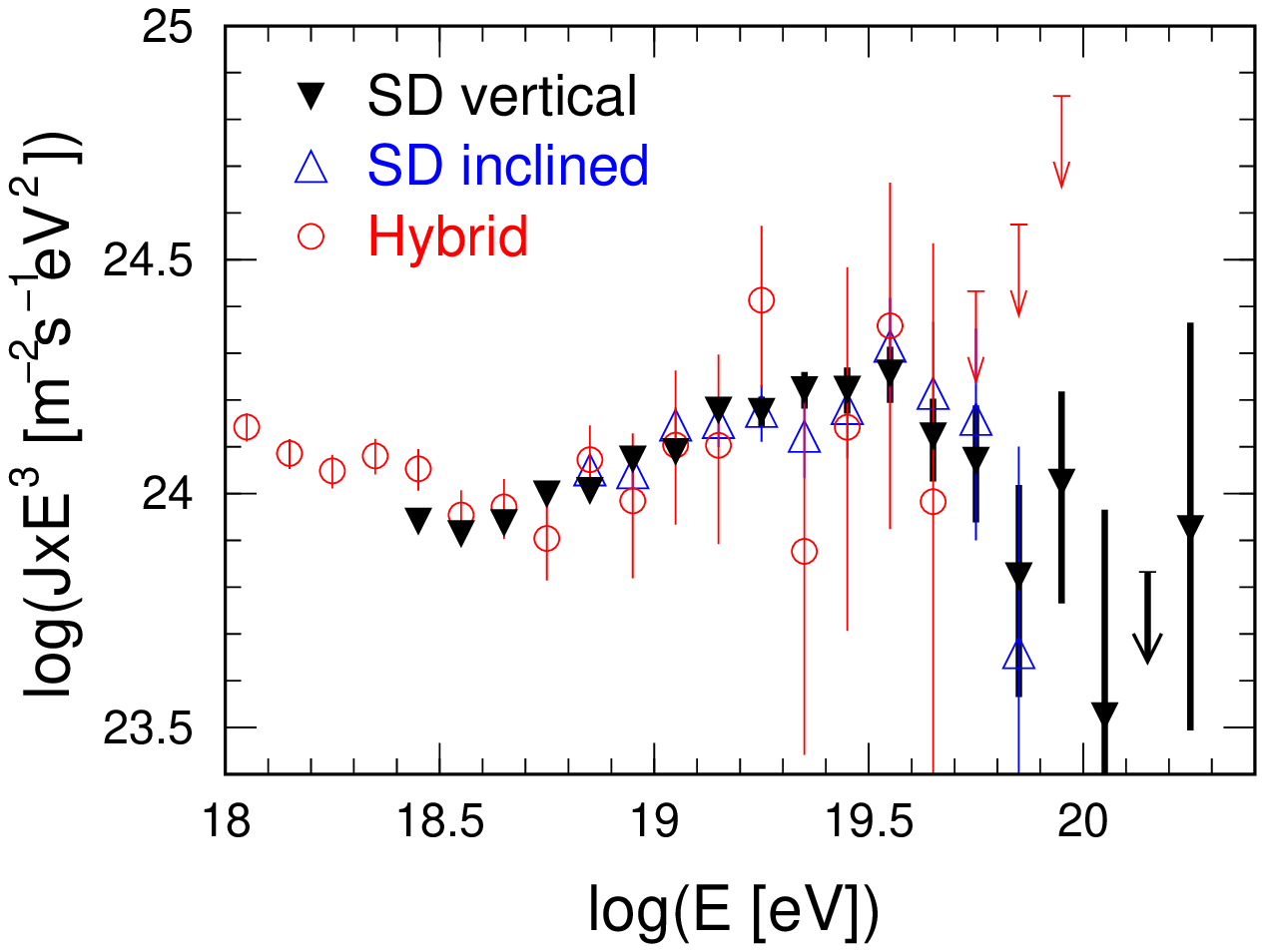}}
\vspace*{-0.5cm} \caption[spectrum]{\raggedright \footnotesize
{Comparison of that spectrum (labeled SD vertical) with the spectra
obtained from hybrid events (Hybrid) and from inclined SD events (SD
inclined), all multiplied by $E^3$ (from
\cite{roth,lorenzo,pedro}).}} \label{fig:spectrumcompa}
\end{minipage}
\label{fig:spectrum}
\end{figure}


The dataset now used to build the spectrum includes all SD T5 events
recorded between January 1st, 2004 and February 28th, 2007, with
reconstructed $\theta \leq 60^\circ$ and energy $E_{FD} \geq 3$ EeV,
which ensures full detection efficiency of the SD and allows a
geometrical computation of the corresponding aperture. After removal
of periods of failure in data acquisition, the corresponding
integrated exposure amounts to about 5165 $\mathrm{km}^2\
\mathrm{sr}\ \mathrm{yr}$, which is more than three times the one
obtained by the AGASA experiment \cite{agasa}. The corresponding
spectrum is shown in Fig.~\ref{fig:spectrumSD} together with its
statistical uncertainty. Although the statistics is still limited,
the hypothesis of a continuation of the UHECR spectrum in the form
of a pure power law beyond $10^{18.6}$ eV can now be rejected at a
$6\sigma$ confidence level, as discussed in \cite{toko}. Efforts to
reduce the systematics in the energy
  estimation are in progress as well. In particular, recent
  and ongoing measurements~\cite{fly} of the fluorescence
yield at a precision level of 5\%
  are expected to significantly improve the accuracy in the reconstructed
  $E_{FD}$.


\subsection{Spectrum from inclined events}

The use of Cherenkov water tanks as surface detectors allows the
Auger to detect showers with zenith angles up to $90^\circ$ (and
even beyond). The range of inclined showers, $60^\circ \leq \theta
\leq 90^\circ$, contributes half the total solid angle of the
detector and about 25\% of its geometrical acceptance, thereby
significantly increasing the field of view of the detector and the
SD statistics.

Such showers are characterized by a dominance of the muonic
component at ground and by a very elongated and asymmetrical
footprint which can exhibit a lobular structure due to the bending
action of the geomagnetic field. The energetic (10 -- 1000 GeV)
muons reach the detector in a thin front with small curvature, which
produces short and peaked FADC pulses. Dedicated selection
procedures and reconstruction methods, based on the use of density
maps of the number of muons at ground, have been developed to
analyze such events; 
more detail can be found in \cite{david}.

The strategy for building a spectrum is the same as in Sec.
\ref{subsec:strategy}. Once the arrival direction is reconstructed,
the pattern of signals is fitted to muon density maps obtained from
simulated proton showers at $10^{19}$ eV in order to determine the
core position and an overall normalisation factor, $N_{19}$, which
acts as an energy estimator and can be calibrated on the FD energy
using inclined hybrid events. No constant intensity cut is needed
because the muon maps already account for the shower attenuation and
geometrical effects. Inclined events with $60^\circ \leq \theta \leq
80^\circ$ and $E_{FD} \geq 6.3$ EeV ($N_{19} \geq 1$), where the SD
detection efficiency is expected to be 100\%, have been used to
build this independant spectrum~\cite{pedro}. The corresponding
integrated exposure amounts to 1510 $\mathrm{km}^2 \mathrm{sr}
\mathrm{yr}$, about a quarter of that of "vertical" ($\theta \leq
60^\circ$) events.

\subsection{Spectrum from hybrid events}

Although their statistics is much smaller, hybrid events 
alone allow a spectrum determination below the energy threshold of
the SD~\cite{lorenzo}. To guarantee the quality of the
reconstruction, only events with a reconstructed $\theta \leq
60^\circ$ and satisfying extra requirements on the observed profile
were selected. In particular, the contamination by Cherenkov light
may not exceed 50\% and the reconstructed depth of the shower
maximum, $X_{max}$, must be observed and lie within a fiducial
volume (which depends on the energy) in order to avoid biases due to
the limited field of view.


The hybrid exposure is estimated on basis of a detailed Monte Carlo
simulation which accounts for the growth of both FD and SD during
the data taking period, as well as for seasonal and instrumental
effects. The hybrid trigger efficiency reaches 100\% at $10^{18}$
eV, independently of the nature of the primary (proton/iron). The
main sources of uncertainty again lie in the determination of the
energy (and its impact on the event selection and aperture
calculation), the knowledge of atmospheric conditions and the
estimation of the detector uptime; see \cite{lorenzo} for a more
detailed discussion.

The three spectra (multiplied by $E^3$) are compared in
Fig.~\ref{fig:spectrumcompa}. The spectrum from SD vertical events
is the most accurate and statistically significant. The hybrid
spectrum extends to lower energies and encompasses the ``ankle",
which appears as a spectral break at $\sim 10^{18.5}$ eV. A detailed
assessment of the sources of systematics remains to be done for
inclined events, but all three spectra are in reasonable agreement
within current estimated uncertainties.

\section{The arrival direction of UHECR: anisotropy searches}
\label{sec:ARG}

Anisotropies in  the flux of UHE cosmic rays may appear in different
energy ranges and angular scales, depending on the nature, distance
and extension of the source(s). Cosmic rays around an EeV are
thought to be of galactic origin, and the region of the Galactic
Center (GC) and the Galactic Plane (GP) are key targets for
anisotropy searches performed with Auger data in that energy range.
At higher energies one rather expects UHE cosmic rays to come from
extra-galactic sources; a search for directional excesses of cosmic
rays could then reveal a correlation with astrophysical objects or
even exotic sources.

The anisotropy studies performed by  Auger are based on both SD T5
and hybrid events with $\theta < 60^\circ$ and the energy assigned
via the cross-calibration procedure described in
Sec.~\ref{subsec:strategy}.

\subsection{Anisotropy studies around the Galactic Center and the Galactic Plane}

In the past, two cosmic ray experiments, AGASA and SUGAR, have
claimed significant excesses in the flux of UHECR in the region of
the GC~\cite{AGASACG,SUGAR}.
Recent TeV $\gamma$-ray observations by HESS~\cite{HESSCG,HESSGP}  
have provided additional hints towards the presence of powerful CR
accelerators in the Galaxy. In that context, several models that
predict a detectable flux of neutrons from the GC in the EeV range
(when the neutron decay length is about the distance from the GC to
the Earth) have also been proposed~\cite{neutronGC}.


With the GC well in the field of view and a much better angular
resolution than previous CR experiments, the Pierre Auger
Observatory is well suited to look for UHECR anisotropies coming
from that region. Such a search was performed on the bulk of SD data
in different energy ranges and with different sizes of the angular
filtering in order to match the resolution of previous experiments.
Using a data sample much larger than the AGASA and SUGAR ones (79265
SD events and 3934 hybrid events with $\theta \leq 60^\circ,
10^{17.9} \ \mathbf{eV} < E < 10^{18.5} \ \mathbf{eV}$,
corresponding to the period from January 2004 to March 2006), Auger
did not confirm any of the anisotropy claims~\cite{AugerCG}.

The same data were also used to search for a point source in the
direction of the GC itself at the scale of Auger's own angular
resolution~\cite{AugerCG}. Applying a $1.5^\circ$ Gaussian filter to
account for the pointing accuracy of the SD, no excess of events was
observed. Assuming both the source and the bulk CR at those energies
have a spectrum index of $3.3$ and that the emitted CRs are protons,
an 95\% C.L. upper limit of $\Phi_s^{95} \leq \xi \ 0.13\
\mathrm{km}^{-2}\ \mathrm{yr}^{-1}$ (where $\xi$ parameterizes the
uncertainties on the flux normalization) is set on the source
flux~\footnote{This bound could however be about $30\%$ higher if
the CR composition at EeV were heavy.}.

A recent update of this analysis with a better angular accuracy and
a significantly larger dataset allowing to split the energy range in
$0.1 \mathrm{EeV} \leq E \leq 1 \mathrm{EeV}$ and $1 \mathrm{EeV}
\leq E \leq 10 \mathrm{EeV}$ have confirmed all negative anisotropy
results and improved the bound on $\Phi_s^{95}$ to $\xi\ 0.018\
\mathrm{km}^{-2}\ \mathrm{yr}^{-1}$. Such a limit already excludes
most of the models of neutron production at the GC~\cite{santos}.

Finally, several methods have been set up to search for large-scale
anisotropies in the distribution of UHECR at energies around the EeV
(and above); such angular patterns would hint towards a galactic
origin of the UHECR just below the ankle. With the current data set,
the right ascension distribution is found to be compatible with an
isotropic sky~\cite{armengaud}. Searches for bidimensional patterns,
such as a possible dipole, are also ongoing.


\subsection{Searches for localized excesses and correlations with astrophysical objects}

Blind searches using Auger data have been performed looking for
small- and intermediate-scale excesses in the sky that would reveal
the presence of point-like sources. The statistical significance of
such an excess is estimated by  calculating the two-point angular
correlation function, which counts the number of pairs of events
with energy larger than a given threshold $E_{th}$ separated by less
than an angle $\theta$. Recent studies on SD T5 data with $E> 10$
EeV, scanning a large range of $(\theta, E_{th})$, shows no really
significant signal of anisotropy, although some hints of clustering
exist at very high energies and intermediate angular
scale~\cite{mollerach}.

Events with energies above 10 EeV have also been used to test a
possible correlation with subsets of BL Lacs, in relation with
previous (and sometimes contradictory) claims and results based on
data from AGASA, Yakutsk and HiRes experiments~\cite{BLLacs}. With 6
times more events than the other existing data samples, the analyzed
Auger data is still compatible with isotropy and does not support
any of the previously reported signals of clustering.

Anisotropy searches based on Auger list of prescribed targets with
definite angular and energy windows as released in
\cite{ICRColdpres} has also given negative results. As more data is
streaming in, the catalogue of candidate targets that will be
studied is expected to increase in the future.

\section{The nature of UHECR: composition studies}
\label{sec:compo}

Thanks to its hybrid design, Auger can in principle measure an extended set of parameters sensitive to the primary UHECR nature and mass. While the discrimination between different types of nuclei is complicated by the uncertainties in the hadronic models, several methods have already been proposed for the identification of photon and neutrinos. The detection of such particles in the UHE cosmic radiation would probe many exotic models of UHECR production and help locate candidate sources as they travel undeflected by the ambient magnetic fields.

\subsection{Upper limit on the flux of UHE photons}
\label{sub:photon} Unlike protons and nuclei, the development of
photon showers is driven by electromagnetic (EM) interactions and
does not suffer much from the uncertainties in hadronic models.
Their development is also
delayed due to the small multiplicity in EM interactions and
 to the LPM effect \cite{LPM},
which reduces the bremsstrahlung and pair production cross-sections
above 10 EeV. 

One of the methods set up by Auger to identify photon primaries in
the flux of UHECR is based on the direct observation of the
longitudinal profile of the shower by the FD; the discriminating
variable is the atmospheric depth of the shower maximum, $X_{max}$
(the estimated average difference in $X_{max}$ between photons and
hadrons is about $200 \mathrm{gr/cm}^2$). The data set used for this
analysis are hybrid events with a reconstructed energy $E > 10^{19}$
eV registered between January 2004 and February 2006. A series of
cuts were applied to guarantee the quality of the hybrid geometry
and of the fit to the shower longitudinal profile (which takes into
account the local atmospheric conditions), and to minimize the bias
against photons introduced in the detector acceptance by requiring
the $X_{max}$ to be inside the field of view.  For all (29) events
passing the cuts, the observed $X_{max}$ is well below the average
value expected from the simulation of 100 photons showers in the
same conditions. Taking systematic uncertainties on the $X_{max}$
determination and the photon shower simulations into account, this
analysis, described in~\cite{Augerphotons}, allowed to put a 95\%
C.L. upper limit on the photon fraction of 16\% above 10 EeV; it has
been recently updated to 13\% using a more extended data
sample~\cite{healy} , as shown in Fig.~\ref{fig:photons}.

Another analysis relying on the SD measurements has also been
developed; the key observables are here the signal risetime at 1000
m ({\it i.e.} the time it takes for the signal to rise from 10\% to
50\%) and the radius of curvature of the shower front. Particles
from showers with a larger $X_{max}$ (and thus a later development)
are indeed expected to reach the ground in  a thicker and more
curved front. A principle component analysis combining both
observables was used to search Auger data for photons; no candidate
was found and the corresponding upper limit on the photon fraction
is 2.0\%, 5.1\% and 31\% at 10, 20 and 40 EeV respectively.

As shown in Fig.~\ref{fig:photons}, the stringent limits put by Auger results
on the UHE photon fraction now disfavour many of the top-down models proposed
in connection with the AGASA spectrum.

\subsection{Upper limit on the flux of UHE neutrinos}

Due to their small interaction cross-section, neutrinos can
penetrate large amounts of matter and generate showers at any
atmospherical depth, unlike protons or photons. Young and deep
neutrino-induced showers can thus be efficiently identified in the
range of inclined showers, $\theta \geq 60^\circ$, by requiring the
presence of a significant EM component.

Upward-going tau neutrinos that graze the Earth just below the
horizon could also be detected as they are likely to interact in the
crust and produce a tau lepton which may emerge and initiate an
observable shower, provided it does not decay too far from the
ground\cite{fargion}. This channel had been pointed out as likely to
increase the detection potential of Auger SD for neutrinos in the
EeV range \cite{nutau}; an extensive study has now been performed on
the available data~\cite{icrcNu1,icrcNu2}.

\begin{figure}[t]
\begin{minipage}[t]{0.45\textwidth}
\mbox{}\\
\hfill \centerline{\includegraphics[width=\textwidth]{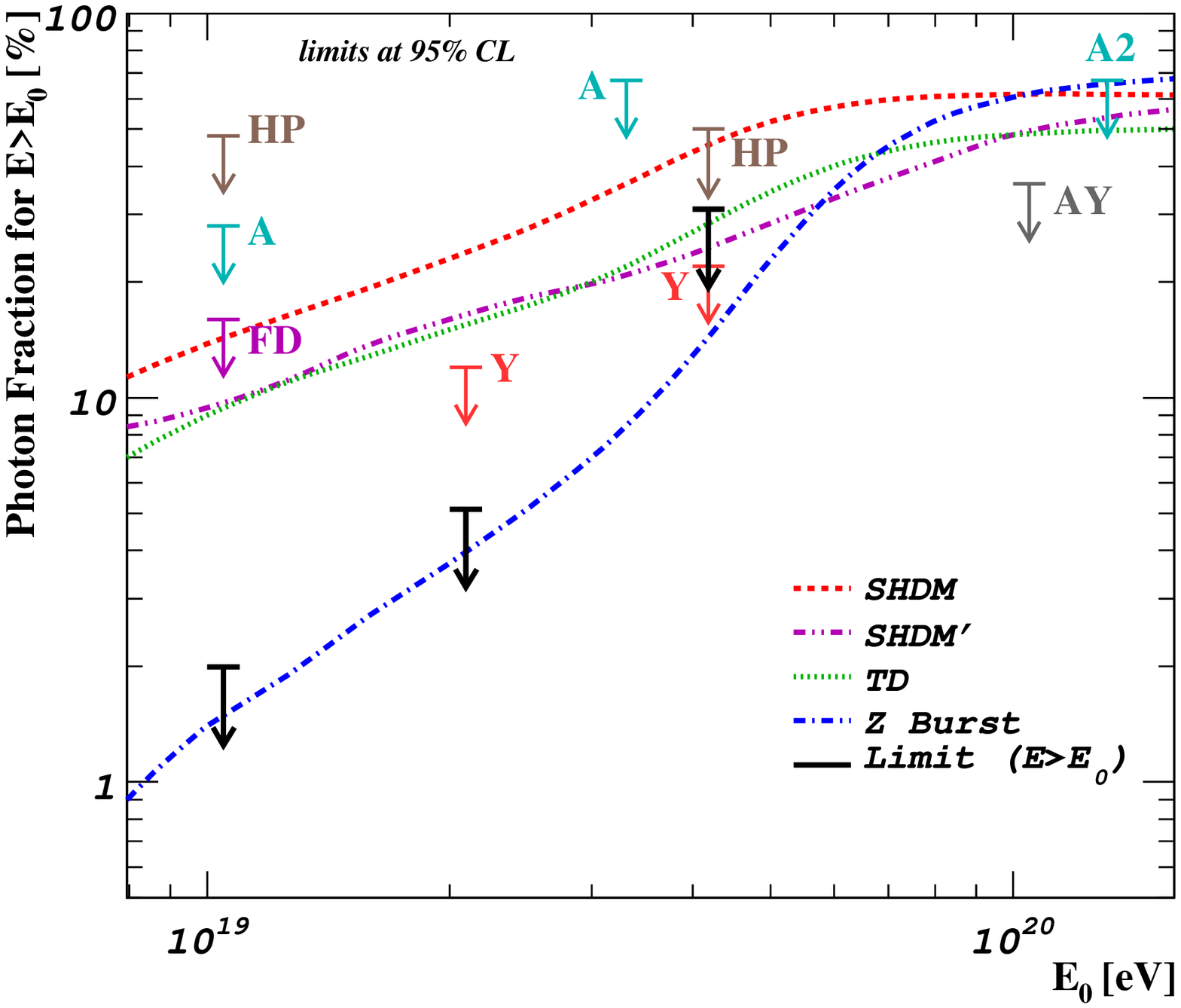}}
\hfill \vspace*{-0.7cm} \caption[Photon]{ \raggedright Auger upper
limits on the UHE photon fraction from the hybrid analysis (labeled
FD) and from SD analysis (black arrows), together with some
predictions from top-down models and the bounds put by previous
experiments (from \cite{healy}).} \label{fig:photons}
\end{minipage}
\hfill
\begin{minipage}[t]{0.45\textwidth}
\mbox{}\\
\hfill \centerline{\includegraphics[width=0.9 \textwidth]{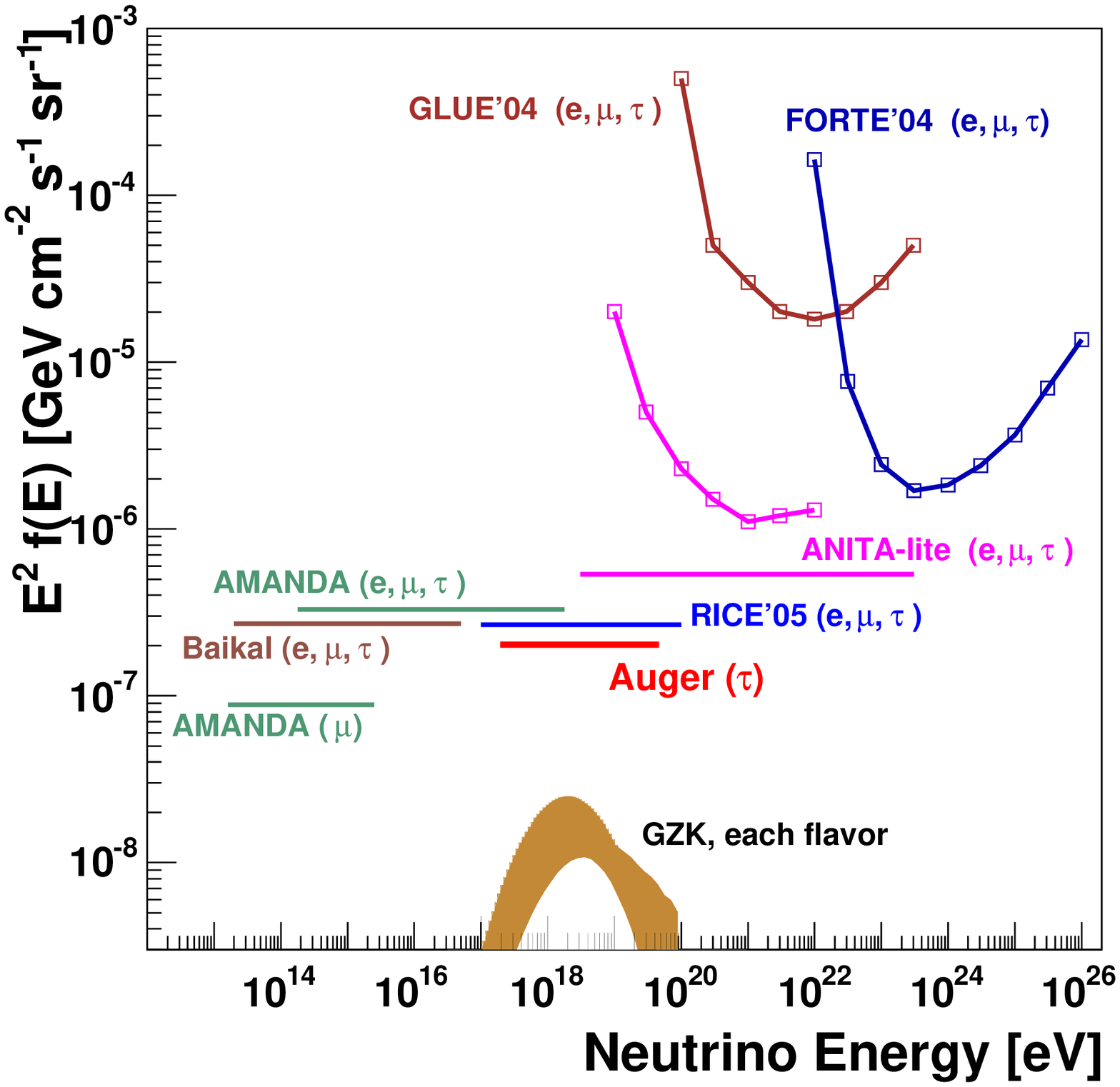}}
\hfill \vspace*{-0.8cm} \caption[NuBound]{ \raggedright Auger upper
limit on an $E^{-2}$ diffuse flux of UHE $\nu_\tau$ (with the worst
estimation of systematics), together with some predictions for GZK
neutrino fluxes and the bounds put by other cosmic neutrino
experiments (from~\cite{icrcNu2}).} \label{fig:nu}
\end{minipage}
\end{figure}

The emerging $\tau$ flux corresponding to a given incident
$\nu_\tau$ flux has been computed in the
relevant angular window 
using both
Monte Carlo and semi-analytical methods accounting
for all  $\nu_\tau \longleftrightarrow \tau$ conversion processes, as well as for the
$\tau$ energy
losses.
The atmospheric shower produced by the decay of the emerging $\tau$ is then
simulated and tested
for detection in the SD.

A specific selection procedure has been set up to identify 
those $\tau$-induced, nearly horizontal showers that develop close
to the detector. The signal shape must be compatible with the
presence of a significant EM component (in practice a ToT-type
trigger 
is required) and a large ($> 1.4$) area-to-peak ratio to reject
triggers produced by consecutive muons. The footprint of those
stations is then required to assume an elongated shape and the
timings to be compatible with a shower front traveling nearly
horizontally at the speed of light. The efficiency of identification
of a $\tau$-induced shower depends on
$E_\tau$ and on $h_c$, 
the altitude of the shower center (defined at a nominal distance of
10 km from the $\tau$ decay point along the shower axis), but also
on the relative position of the footprint in the array.  To compute
the detector acceptance for $\nu_\tau$, a double Monte Carlo
integration accounting for the evolution of the array with time is
performed.

SD data from January 2004 till December 2006 were searched for
candidate grazing $\nu_\tau$'s, but no single event passed the
selection criteria, which allows to derive an upper limit for any
injected flux of UHE $\nu_\tau$ with a given shape. Assuming an
$E^{-2}$ incident spectrum of diffuse $\nu_\tau$, a 90\% C.L. bound
$E_\nu^2 \dot dN_{\nu}/dE_\nu < 1.5^{+0.5}_{-0.8}\ 10^{-7}\
\mathrm{GeV}\ \mathrm{cm}^{-2}\mathrm{sr}^{-1}\mathrm{s}^{-1}$ was
derived in the energy range $[2\,10^{17}-\,5\,10^{19}]$ eV. The
sources of systematic uncertainties have been carefully
addressed~\cite{icrcNu2}; they are globally responsible for a factor
of $\sim 3$ uncertainty on the acceptance, which propagates to the
final flux limit. Among them, physical quantities that have not been
measured at those energies, such as the $\nu$ cross-sections, the
$\tau$ polarization and energy losses, contribute resp. $\sim 15\%$,
$\sim 30\%$ and $\sim 40\%$ . The Monte Carlo simulations of the
shower and the detector response add an extra $\sim 25\%$
uncertainty, and the effect of neglecting the actual topology of the
Auger site 
another $\sim 18\%$. As shown in Fig.\ref{fig:nu}, Auger limit is
nevertheless the best to date in the energy range where GZK neutrino
fluxes (produced by the interaction of the observed UHECR with the
cosmic microwave background) are likely to peak. To improve that
limit by an order of magnitude or so will however require the
accumulation of several more years of data.


\section{Conclusions}
The past three years have witnessed a phase of major development of
the Southern Auger Observatory on the field, accompanied by a
significant increase of the dataset. 
A lot of progress has been made in the understanding the detector,
which resulted in a better control on the systematic uncertainties
an in the development of reliable and robust analysis methods which
allowed the release of first scientific results concerning the UHECR
spectrum and angular distribution. If a continuation of the spectrum
above ${10}^{20}$ eV seems unlikely, a much larger data sample is
still needed to determine the exact shape of the spectrum. Auger
also sees a remarkably isotropic sky, except maybe at high energies
where more data are necessary for a detailed study of clusterings
and correlations, whatever the scale.
 Finally, Auger has already put competitive limits on the fluxes of UHE photons
and neutrinos, thereby demonstrating its capabilities to work as a
multi-messenger detector.
\vspace*{-0.1cm}
\footnotesize{
\section*{Acknowledgments}

The author wishes to thank both the organizers, for making so many
fruitful exchanges between theorists and experimentalists possible,
and the European Community $\mathrm{6^{th}}$ F.P. for supporting its
activities in Auger through the Marie Curie Fellowship MEIF-CT-2005
025057. }


\vspace*{-0.1cm}
\section*{References}
\begin{thebibliography}{99}
  \bibitem{agasa} M. Takeda {\it et al.}, \APP 19 (2003) 447.
  \bibitem{HiRes}  T.~Abu-Zayyad {\it et al.}  [HiRes Collaboration],
  {\em Astropart.\ Phys.} 23 (2005) 157.
  \bibitem{phys} A.~M.~Hillas,
  arXiv:astro-ph/0607109; D.~F.~Torres and L.~A.~Anchordoqui,
  {\em Rept.\ Prog.\ Phys.} 67 (2004) 1663; 
  P.~Bhattacharjee and G.~Sigl,
  {\em Phys.\ Rept.} {\bf 327}, 109 (2000).
\bibitem{opa} J.~Abraham {\it et al.}  [Pierre Auger Collaboration],
  Nucl.\ Instrum.\ Meth.\  A {\bf 523} (2004) 50.
\bibitem{triggers} D. Allard {\it et al.} \icrcold, 287.
\bibitem{carla} C.~Bonifazi {\it et al.}  [Pierre Auger
                  Collaboration], 
arXiv:0705.1856 [astro-ph].
\bibitem{sommers} P. Sommers \icrcold, 387.
\bibitem{ave} M. Ave \icrc
\bibitem{dawson} B.~R.~Dawson  [Pierre Auger Collaboration],
  arXiv:0706.1105 [astro-ph].
\bibitem{roth} M.~Roth [Pierre Auger Collaboration],
  arXiv:0706.2096 [astro-ph].
\bibitem{toko} T. Yamamoto \icrc
\bibitem{fly} 
G.~Lefeuvre {\em et al.},
arXiv:0704.1532 [astro-ph]; F.~Arciprete {\it et al.},
  {\em Nucl.\ Phys.\ Proc.\ Suppl.} 150 (2006) 186.
\bibitem{lorenzo} L. Perrone [Pierre Auger Collaboration], arXiv:0706.2643 [astro-ph].
\bibitem{david} D. Newton \icrc
\bibitem{pedro} P. Facal San Luis \icrc
\bibitem{AGASACG} N.~Hayashida {\it et al.}  [AGASA Collaboration],
  Astropart.\ Phys.\  {\bf 10} (1999) 303
 ; M. Teshima {\it et al.} [AGASA Collaboration], {\em Proc. 27th
 ICRC} 1 (2001) 337.
\bibitem{SUGAR} J.~A.~Bellido {\it et al.}, 
  {\em Astropart.\ Phys.\ } 15 (2001) 167.
\bibitem{HESSCG} F.~Aharonian {\it et al.}  [HESS Collaboration],
  {\em Astron.\ Astrophys.\ }  425 (2004) L13.
\bibitem{HESSGP} F.~Aharonian {\it et al.}  [HESS Collaboration],
  {\em Nature} 439 (2006) 695.
\bibitem{neutronGC} see e.g. G. Medina Tanco, A. Watson, {\em Proc. 27th ICRC} (2001)
531; R. Crocker {\it et al.}, {\em Astrophys. J.} 622 (2005) 273; F.
Aharonian, A. Neronov, {\em astrophys. J.} 619 (2005) 306.
\bibitem{AugerCG} M.~Aglietta {\it et al.}  [Pierre Auger Collaboration],
  {\em Astropart.\ Phys.\ } 27 (2007) 244.
  \bibitem{santos} E.M. Santos [Pierre Auger Collaboration], arXiv:0706.2669 [astro-ph].
  \bibitem{armengaud} E.~Armengaud [Pierre Auger Collaboration],
  arXiv:0706.2640 [astro-ph].
  \bibitem{mollerach} S. Mollerach [Pierre Auger Collaboration], arXiv:0706.1749 [astro-ph].
  \bibitem{BLLacs} P. Tinyakov and I. Tkatchev, {\em JETP Lett.}
  74 (2001) 445; D. Gorbunov {\it et al.}, {\em JETP Lett.}
  80 (2004) 145; R. Abbasi {\it et al.} [HiRes Collaboration], {\em Astrophys.
  J.} 636 (2006) 680.
  \bibitem{ICRColdpres} R. Clay [Pierre Auger Collaboration], {\em Proc. 28th ICRC} 1 (2003) 421.
\bibitem{LPM} L. D. Landau, I. Ya. Pomeranchuk Dokl. Akad. Nausk. SSSR {\bf 92} (1953), 535 \& 735; A. B. Migdal, Phys. Rev. {\bf 103} (1956), 1811.
\bibitem{Augerphotons} J.~Abraham {\it et al.}  [Pierre Auger Collaboration],
  {\em Astropart.\ Phys.\ } 27 (2007) 155.
\bibitem{healy} M.D. Healy \icrc
\bibitem{fargion} D. Fargion, {\em Astrophys.\ J.}\ 570 (2002) 909 and references therein.
\bibitem{nutau} 
X. Bertou {\it et al.}, \APP 17 (2002) 183.
\bibitem{icrcNu1} J. Alvarez-Mu\~{n}iz, \icrc
\bibitem{icrcNu2} O.~B.~Bigas [Pierre Auger Collaboration],
  arXiv:0706.1658 [astro-ph].

\end{thebibliography}
\end{document}